\documentstyle[aps,twocolumn,graphics,epsf,amsmath,amssymb,eqsecnum,floats]{revtex}
\begin{document}

\title{ 1-Dimensional behavior and sliding Luttinger liquid phase
in a frustrated spin-1/2 crossed chain model:  
contribution of exact diagonalizations.
}

\author{P. Sindzingre, J.-B. Fouet, C. Lhuillier}
\address{Laboratoire de Physique Th{\'e}orique des Liquides-UMR 7600
of CNRS,  Universit{\'e}  Pierre  et  Marie  Curie,  case 121,   4 place
Jussieu, 75252 Paris Cedex, France\\ E-mail:phsi@lptl.jussieu.fr}
\maketitle
\bibliographystyle{prsty}
(\today)\\
\begin{abstract}
Exact diagonalizations indicate that the  effective 1-dimensional
behavior (sliding Luttinger liquid phase)
of the frustrated spin-1/2 crossed chain model,
predicted by Starykh, Singh and Levine 
[Phys. Rev. Lett. {\bf 88}, 167203 (2002)],
persists for a  wide range  of transverse couplings.
The extension of  the other phases (plaquette valence bond 
crystal and N\'eel long range order) is precised.
No clear indication of a coexistence of these two phases is found,
at variance with a suggestion of Sachdev and Park (cond-mat/0108214).
\end{abstract}

\section{INTRODUCTION}
\begin{figure}
        
\hspace*{0.2cm} 
\resizebox{4cm}{!}{\includegraphics{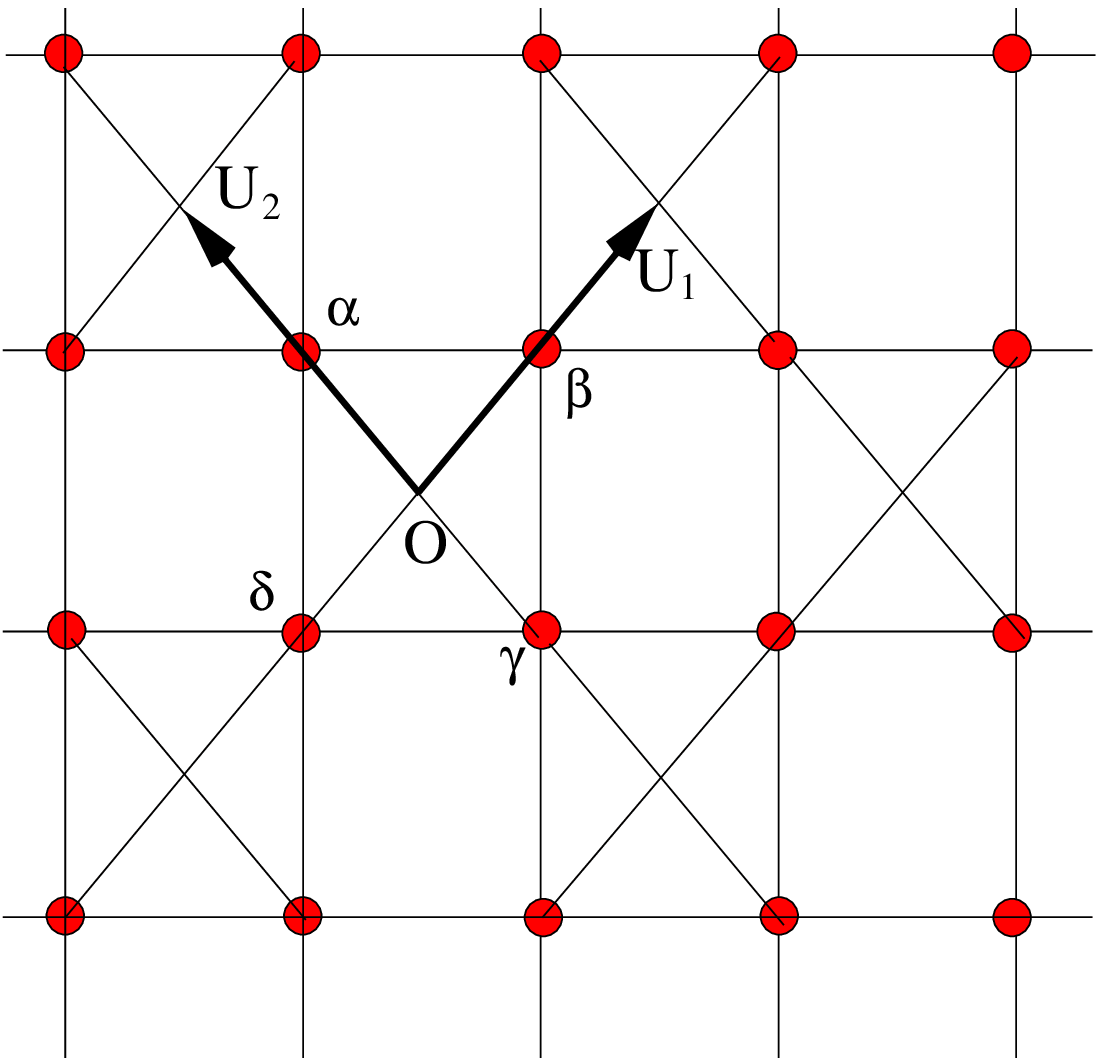}}\\

\vspace*{-4.15cm}
\hspace*{4.7cm} \resizebox{3.5cm}{!}{\includegraphics{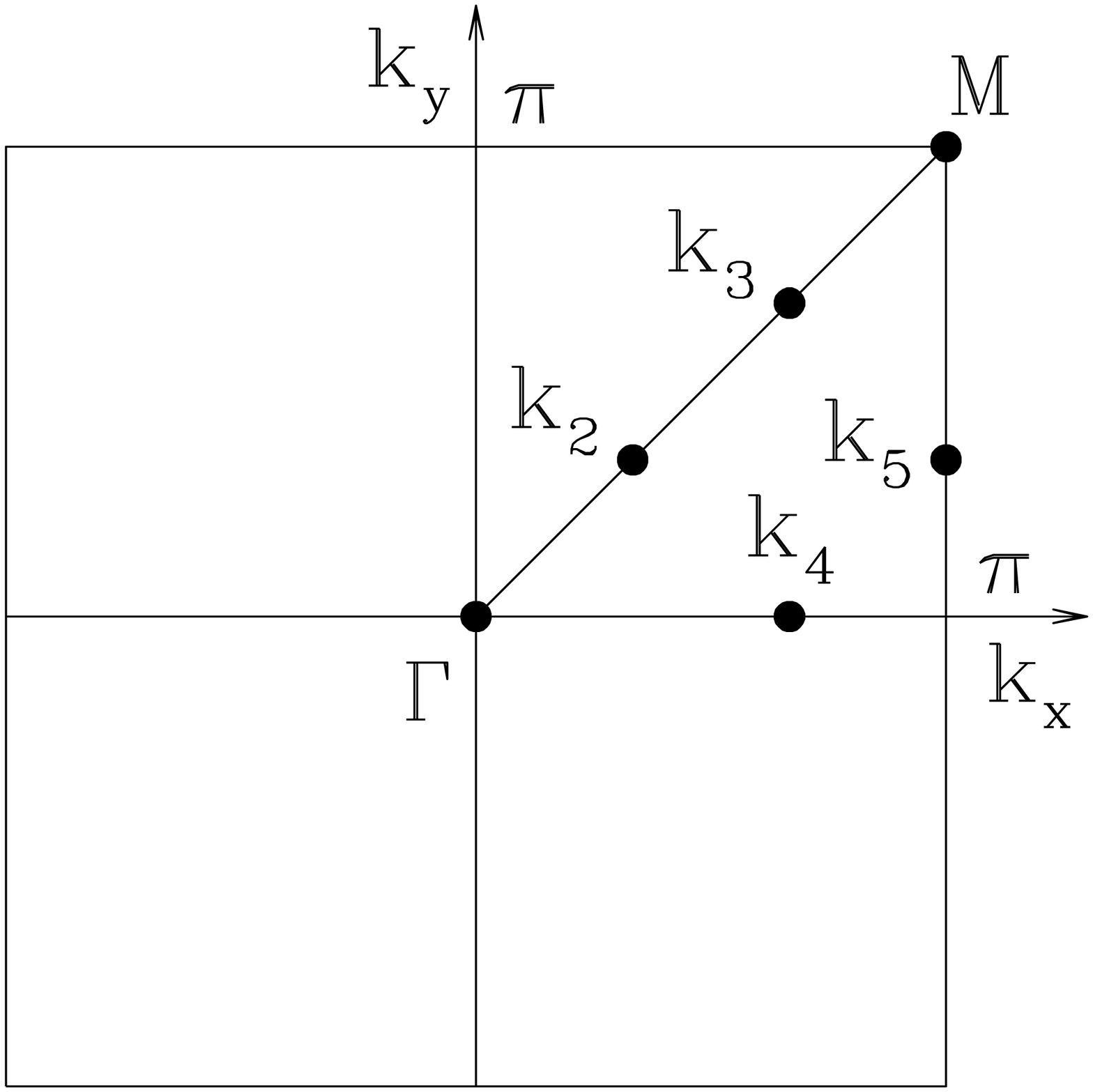}}\\

\vspace*{0.4cm}

        \caption[99]{  Left: the crossed chain magnet. 
The spins sit at the vertices shown by bullets, 
the coupling constant between nearest-neighbor pairs is $J_1$,
the coupling constant on the diagonal bonds of the checkerboard of plaquettes
is $J_2$.
$\bf u_1, u_2$ are the unit vectors of the Bravais lattice.
Right: the Brillouin zone associated to this lattice. $\Gamma$, 
 $k_i \; (i=2,5)$,  M are the  six different
wave vectors present in the N=36 sample.
        }  \label{checkerboard2}
\end{figure}

Among the family of frustrated two dimensional  spin-1/2 
antiferromagnetic (AF) models, a crossed chain model (CCM), 
which may  also be viewed as an extension 
of the  checkerboard  antiferromagnet, 
has recently attracted interest in two distinct ranges of 
parameter\cite{ssl02,ssf98,c01,fmsl01}.
The CCM Hamiltonian reads:
\begin{equation}
{\cal H} =  J_1 \sum_{<i,j>} {\bf S}_i.{\bf S}_j
\,+\,  J_2 \sum_{<<i,j>>} {\bf S}_i.{\bf S}_j
\label{eq-Heis}
\end{equation}
where the exchange $J_1$ couple nearest-neighbor pairs $<i,j>$ of spins
on a square lattice and exchange $J_2$ next-nearest-neighbor pairs $<<i,j>>$
on a checkerboard pattern of plaquettes as shown in  Fig.~\ref{checkerboard2}.
Alternatively, $J_2$ may be viewed as the intra-chain exchange between
spins on the diagonal chains  and $J_1$ as the coupling constant between these
crossed chains.
Both $J_1$ and $J_2$ are assumed positive, describing AF couplings.
This model interpolates between the AF Heisenberg model on the square lattice 
% (SAFHM) 
for $J_2=0$ and decoupled AF Heisenberg chains  % (HAFC) 
for $J_1=0$. 
When $J_1=J_2$, one has the checkerboard antiferromagnet which is a 2d
analog of the pyrochlore antiferromagnet.

The physics of the spin-1/2 model is well established at three points.
For $J_2=0$ it has collinear N\'eel long range order (LRO)
with gapless $\Delta S^z = 1$ elementary excitations (magnons).
At $J_2=J_1$ it is a valence bond crystal (VBC) with LRO in
singlet plaquettes \cite{fmsl01}; its elementary excitations are 
gapped and carry integer spins \cite{bh01}.
 When $J_2/J_1 = \infty$, one has the physics of
decoupled AF Heisenberg chains, their elementary excitations are 
gapless spin-1/2 deconfined spinons.

Recently, motivated by experimental results on\\
$Cs_2CuCl_4$~\cite{cttt01} and $Na_2Ti_2Sb_2O$~\cite{ax97},
Starykh, Singh and Levine (SSL) have investigated the CCM
in the limit $J_2/J_1  \gg 1 $ where it describes  "weakly" coupled
 chains~\cite{ssl02}.
They concluded that,
the chains behave then as quasi-decoupled, realizing a $SU(2)$ 
"sliding Luttinger phase" \cite{mkl01,vc01}.
Usually the 1d AF Heisenberg behavior 
is unstable to non frustrating 2d couplings
giving rise to N\'eel LRO or VBC LRO.
SSL argue that in the CCM the 1d behavior should survive for finite 
non negligible $J_1/J_2$ due to the frustrating nature of the
transverse coupling.
 Since the pioneering works of Majumdar-Gosh and Haldane,
 many ladder models  have been studied 
 giving birth to various 1d behaviors\cite{wa96,nge98,ahllt98}.
The case of the spatially anisotropic couplings of the triangular 
lattice typical
of $Cs_2CuCl_4$, appears in some aspects rather exceptional with
non negligible transverse couplings and nevertheless an effective 1d
behavior with a continuum of gapless excitations looking very much
 like Fadeev spinons \cite{be01}.

The first purpose of this paper is to investigate SSL prediction
using exact diagonalizations (ED)  and try to figure out
the possible extension of the quasi-1d regime with decreasing $J_2/J_1$.
The second purpose will be to study the range of stability of the VBC phase
around $J_2/J_1=1$.
% with respect to the anisotropy of couplings  with either increasing
% or decreasing $J_2/J_1$. 
We shall also address  a new
suggestion of Sachdev and Park \cite{sp01}
on the possible coexistence of phases with two different order
parameters between the pure VBC and N\'eel phases.

\section{Results}

The ED calculations have  been carried out on $N=16,32,36$ samples
which display all the symmetries of the infinite lattice, using
periodic boundary conditions as described in \cite{fmsl01}.
Additional calculations using twisted boundary conditions 
were also performed on the $N=16$ sample  
( specifically for $J_2/J_1 > 1$)
which indicated that the lowest  ground-state energies are obtained for
periodic boundary conditions:  it might be inferred from these results
that incommensurate phases are unlikely 
in the absence of a magnetic field \cite{ssl02}.

\begin{figure} [h]
        \begin{center}
        \resizebox{6cm}{5cm}{\includegraphics{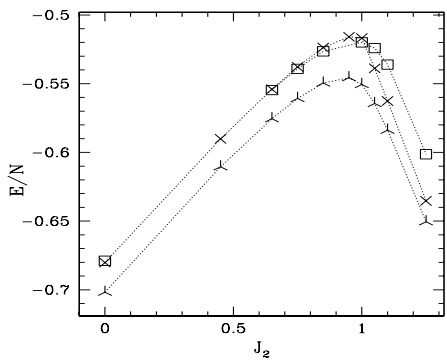}} \end{center}

        \caption[99]{ Ground-state energies per spin $E/N$ vs $J_2$ 
($J_1=1$) for $J_2 \le 1.25$: $N=16$ (tripodes), 
$N=32$ (crosses) and $N=36$ (full squares).
The dotted lines  are guides for the eye.
%shown for $J2>1.25$ depict the values of the
% energies of completely decoupled chains of length 4 (full line) and
% 6 (long-dashed line).
        }  \label{energies}
\end{figure}

\begin{figure} [h]
        \begin{center}
        \resizebox{6cm}{5cm}{\includegraphics{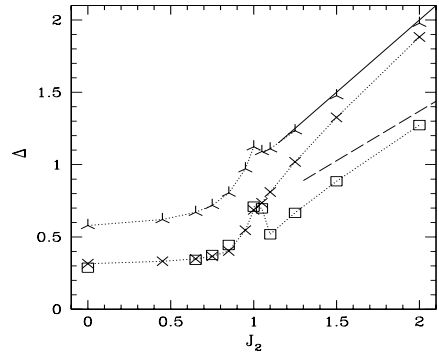}} \end{center}

        \caption[99]{ Spin gap $\Delta$ vs $J_2$ ($J_1=1$), 
same symbols as in  Fig.~\ref{energies}. 
 Extra lines show 
the spin gap values for independent chains of length $L=4$ (full line)
and  $L=6$ (dashed line) for $J2>1.25$.
        }  \label{spin_gaps}
\end{figure}

\begin{figure} [h!]
        \begin{center}
        \resizebox{6cm}{5cm}{\includegraphics{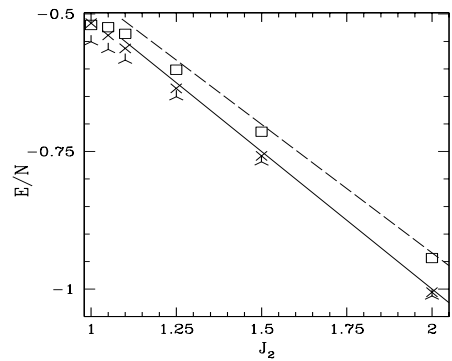}} \end{center}

        \caption[99]{ Ground-state energies per spin 
$E/N$ vs $J_2$ for $J_2 \ge 1$, same symbols as in Fig.~\ref{energies}. 
 Extra lines show 
the $E/N$ values for independent chains of length $L=4$ (full line)
and  $L=6$ (dashed line) for $J2>1.25$.

%shown for $J2>1.25$ depict the values of the
% energies of completely decoupled chains of length 4 (full line) and
% 6 (long-dashed line).
        }  \label{energies_high}
\end{figure}

The general evolutions, as a function of $J_2$ ($J_1=1$), 
of the ground-state energies per spin $E/N$ and the spin gaps $\Delta$ 
are displayed in Figs.~\ref{energies}, ~\ref{spin_gaps} 
and ~\ref{energies_high}.
As shown in Fig.~\ref{energies} the point of "maximum frustration" where 
$E/N$ has its maximum, is reached for $J_2/J_1$ slightly smaller than one,
close to the checkerboard point. 
On might notice that the finite size evolutions between N=32 and
N=36 are hardly visible on the scale of these figures for $J_2/J_1 <1$ 
but are dramatic for $J_2/J_1>1$. We will show below that
this last behavior is related to the quasi- 1-dimensional
behavior of the model in this range of parameter.
%We consider first the region on the right of this point to assess the
%extension of 1d behavior.   

\subsection{ Quasi 1d behavior}
ED results show a large extension of 1d behavior 
from $J_2/J_1=\infty$ down to $J_2/J_1=1.5$
and probably below.

First, Fig.~\ref{energies_high} %and Fig.~\ref{spin_gaps}
shows that $E/N$  is quite close to the
energy values of decoupled chains of same length  for 
$J_2/J_1 \ge  1.10 $.
 Comparison of results for N=16
and N=32 shows that increasing the number of chains pushes the
energy per spin toward the value of uncoupled chains 
(Note that the N=16 sample has $2\times 2$ chains of length 4, 
the N=32 sample has $2\times 4$ chains of length 4, 
whereas the N=36 sample has $2\times3$ chains of length 6).
 Pointing to the same conclusion,
the spin gap (Fig.~\ref{spin_gaps}) shows dramatic finite size
effects between N=32 and N=36 for $J_2/J_1 \ge  1.10 $.

 Second, 1d behavior down to at least $J_2/J_1=1.5$ 
is supported by spin-spin correlations:
\begin{equation} 
s(ij)=< {\bf S}_i.{\bf S}_j >
\end{equation}
and dimer-dimer correlations patterns:  
\begin{equation}
 D(ij;kl)=< {\bf S}_i.{\bf S}_j  {\bf S}_k.{\bf S}_l >
-  < {\bf S}_i.{\bf S}_j > <  {\bf S}_k.{\bf S}_l > \; .
\label{def-cor-dimer-dimer}
\end{equation}
Values of $s(ij)$ computed in the ground-state of the $N=36$ sample 
at $J_2/J_1=1.5$ are displayed in Table~\ref{tab-cor-spin-spin}. 

\begin{table}[h!]
\begin{center}
\begin{tabular}{|c|c||c|c|}
  j &   s(ij) &  j  &      s(ij)         \\
\hline
 36 & -0.4551363204      & 4 & -0.0228887132  \\
 15 &  0.1970987646      & 2 & -0.0156898505  \\
 22 & -0.2193602830      & 9 &  0.0102951567  \\\cline{1-2}
  3 &  0.0017845218      &16 &  0.0029547330  \\\cline{3-4}
 10 &  0.0006638989      &17 & -0.0040816720  \\
 12 & -0.0009191508      &   &                \\ 
\end{tabular}
\end{center}
% 36 & -0.4551363204      & 4 & -0.0228887132     \\
% 15 &  0.1970987646      & 2 & -0.0156898505       \\
% 22 & -0.2193602830      & 9 &  0.0102951567     \\\hline\hline
% 3 &  0.0017845218      &12 & -0.0009191508    \\
%10 &  0.0006638989    && \\\hline \hline
%17 & -0.0040816720      &16 &  0.0029547330     \\
\caption[99]{ Spin-spin correlations $s(ij)=\, <{\bf S}_i.{\bf S}_j >$
between site $i=1$ and site $j$
in the exact ground-state of the $N=36$ sample for $J_2/J_1=1.5$.
The sites are numbered as in Fig.~\ref{dimer_cor_j2.1.50}. 
Other correlations can be deduced using  $C_{4v}$ symmetries.
Correlations along a $J_2$ chain 
are displayed in the upper left corner of the
table , those between orthogonal chains 
% associated to paths with only one weak bond $J_1$ 
are in the upper right corner, those between parallel chains in the
lower part.
% Paths between site 1,9  ( 3, 10 and 12)
% involve  2 weak bonds. The pairs of sites (1,17) and
% (1,16) are over-correlated, they can be reached by different paths
% with either three or four  lattice steps. 
}
\label{tab-cor-spin-spin}
\end{table}
They are largest  and decrease slowly  on a line of diagonal bonds and
are quite small if sites $i$ and $j$ belong to different chains, even
if $i$ and $j$ are first neighbors.
Spins on parallel chains are less correlated than those on orthogonal
chains which may be understood in a perturbative approach:
coupling between orthogonal chains
involves one "$J_1$" term instead of two for parallel chains.

%  As can be seen in
%  Table~\ref{tab-cor-spin-spin} the order of magnitude of these
% correlations decrease steadily with  the number of steps on weak
% bonds ($J_1$ bonds).

\begin{figure}
        \begin{center}
        \resizebox{6cm}{!}{\includegraphics{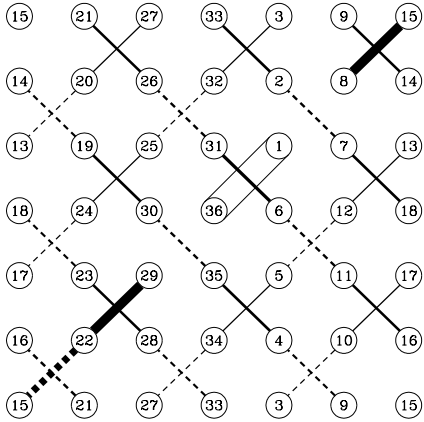}} \end{center}
        \caption[99]{ % Labels of the sites for $N=36$.
Dimer-dimer correlations $D(ij;kl)$ 
in the exact ground-state of the $N=36$ sample ($J_2/J_1=1.5$)
between reference pair $(1,36)$ and pair $(k,l)$.
Full ($D>0$) or dashed ($D<0)$) lines have width proportional
to the square root of $|D|$ (a choice done to ease visualization).
        }  \label{dimer_cor_j2.1.50}
\end{figure}

The behavior of the dimer-dimer correlations is identical 
( Table~\ref{tab-cor-dimer-dimer} and
Fig.~\ref{dimer_cor_j2.1.50}).
$D(ij;kl)$ values between the pair ($i ,j$) on a diagonal bond and pairs $(k,l)$
are also largest if $(i,j)$ and $(k,l)$ are on the same chain. They
decrease by an order of magnitude on nearby regions of crossed
chains and by two orders of magnitude on parallel chains.

\begin{table} 
\begin{center}
\begin{tabular}{|c|c||c|c|}
 (k,l) &   D           &(k,l)  &     D    \\
\hline
 29 22 &  0.0897246082 &34 27 & -0.0015374269 \\ 
 22 15 & -0.0523861622 &10  3 & -0.0015374269 \\\cline{3-4}
 31  6 &  0.0110558121 &34 29 & -0.0011366550 \\ 
 35  4 &  0.0074567919 &27 22 &  0.0005263317 \\
 28 23 &  0.0066321724 &15 10 &  0.0005263317 \\
 16 11 &  0.0066321724 &11  4 & -0.0003686058 \\
 21 16 & -0.0063063310 &35  6 &  0.0001626070 \\
 35 30 & -0.0057705738 &16  9 &  0.0001126614 \\
 11  6 & -0.0057705738 &28 21 &  0.0001126614 \\
  9  4 & -0.0046514788 &35 28 & -0.0000762647 \\
 17 10 &  0.0017481736 &33  4 & -0.0000456425 \\
 34  5 &  0.0016564513 &34  3 &  0.0000268261 \\
 12  5 & -0.0015704399 &10  5 &  0.0000036528 \\
\end{tabular}
\end{center}
\caption[99]{ Dimer-dimer correlations
$D(ij;kl)$ (Eq.~\ref{def-cor-dimer-dimer})
between the pair of sites $(i,j)=(1,36)$ and pairs of sites $(k,l)$
in the exact ground-state of the $N=36$ sample ($J_2=1.5$, $J_1=1$).
The values of $D$ for pairs $(k,l)$ on diagonal bonds 
coupled with $J_2$
(left column and upper right corner above the line
and Fig.~\ref{dimer_cor_j2.1.50})
are all larger than on non-diagonal bonds.
The sites are numbered as in Fig.~\ref{dimer_cor_j2.1.50}.
}
\label{tab-cor-dimer-dimer}
\end{table}

\begin{figure} [h!]
        \begin{center}
        \resizebox{6cm}{5cm}{\includegraphics{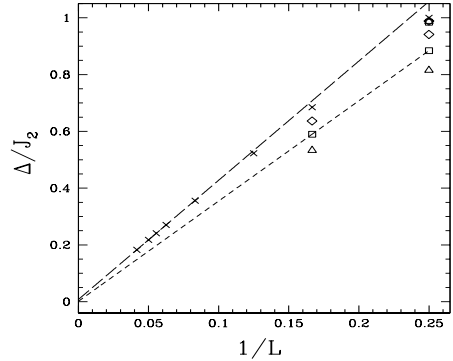}} \end{center}
        \caption[99]{ 
Ratios $\Delta/J_2$  of the spin-gaps to the intra-chain coupling $J_2$ 
of the chains 
at $J_2/J_1=2$ (open losanges),
$J_2/J_1=1.5$ (open squares)  
and $J_2/J_1=1.25$ (open triangles). 
The spin-gaps of the AF Heisenberg chain are shown by crosses.
% which have been fitted to the
%dashed line for $L>14$. 
%At $J_2/J_1=5$ the values of $\Delta/J_2$ (not shown) would be
%indistinguishable  on this figure from the Heisenberg chain values.
All $N=16$ values are very close to the $L=4$ value of the Heisenberg chain.
The dotted line joins the $N=32$ and $N=36$ results for $J_2/J_1=1.5$.
        }  \label{compare_gaps}
\end{figure}

Third, the evolution of the spin-gap $\Delta$ with increasing size,
is not inconsistent with a vanishing value   for $N\rightarrow\infty$,
% far away from the limit $J_2/J_1=\infty$ and probably 
at least down to $J_2/J_1=1.5$.
In Fig.~\ref{compare_gaps}  we display the
values of $\Delta/J_2$ vs $1/L$  
for $N=16,32$ ($L=4$), $N=36$ ($L=6$) and compare them
with the spin-gaps of the AF Heisenberg chain. %  HAFC.
At $J_2/J_1=5$ the values of $\Delta/J_2$ (not shown)
would be indistinguishable from the Heisenberg chain % HAFC 
values and only slightly
deviate from the latter down to $J_2/J_1=1.5$. 
The extrapolation for $L\rightarrow\infty$ at $J_2/J_1=1.5$ is 
 problematic due to  system sizes: nevertheless a linear extrapolation
 of the two sets of data for  $N=32$ and $N=36$ points to a zero gap:
 looking to the sign of the deviation from the 1/L asymptotic limit 
(which is negative) one is indulged to conclude that the system is
 still gapless in the thermodynamic limit for $J_2/J_1=1.5$.
This could even hold at $J_2/J_1=1.25$, 
but the extrapolation is then more uncertain with present sizes.
 
% results are rather well aligned with the origin.
%Of course these  sizes are likely too small to justify a $\sim 1/L$ law
%as this  asymtotic regime  only appear for much larger $L$ in the HAFC. 
%But this suggest that $\Delta$ is at most very small in the
%thermodynamic limit.

   Fourth, a lengthy but straightforward calculation shows that the quantum
numbers of the ground-state and first excitations of the CCM
model for  $J_2/J_1 =5 $ (Table~\ref{ener_j2_5}) are directly
related to the quantum numbers of the ground-state and first
excitations of chains of length $L=4$ or 6 (respectively for the
$N=32$ and 36 samples). The effect of the
transverse inter-chain coupling may be considered in a first
approach as  a minor quantitative effect.
The lowest S=1 excitations 
consist of a set of 4 (N=16), 8 (N=32),
respectively 6 (N=36) quasi degenerate states.
 Their parentage to the first excitation (labelled a in Table~\ref{ener_j2_5})
of the unique chain can be exactly traced back. Their wave vectors 
are vectors of the side of the square Brillouin zone
(points $k_5$ and M of the Brillouin zone in Fig.~\ref{checkerboard2}
for $N=36$) .
Appropriate superpositions of these quasi degenerate states 
create S=1 excitations which are essentially localized on one
given chain!

\begin{figure}  [h!]
\begin{center}
\resizebox{6cm}{5cm}{\includegraphics{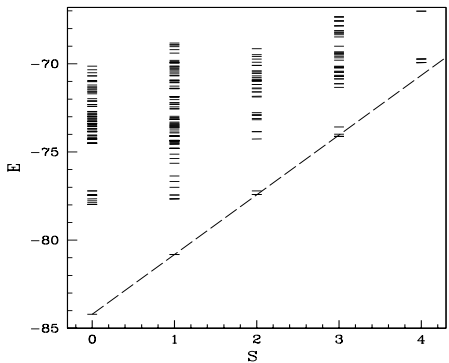}} \end{center}
\caption[99]{    Spectrum of eigen-energies $E$ in each total
       sector of spin $S$ for $N=36$ at $J_2/J_1=5$.
This spectrum is quite close to the uncoupled chains one (see
Table~\ref{ener_j2_5} ).
The lowest energies $E(S)$ in each $S$ sector increase linearly with $S$ up to
$S=3$ as shown by the dashed line.
        }  \label{spec_s_sqm36_5.00}
\end{figure}

 The quasi degenerate states with total energy $\sim 77-78$ 
in Fig.~\ref{spec_s_sqm36_5.00} are associated to the creation
of excitations with 2 ``a-quanta'' on different chains
 (levels 2a in Table~\ref{ener_j2_5}).
This process gives birth to states with total spin $S=0, \,1,\,2$.
Endly we should notice that the linear increase
with S of the lowest eigen-states in each spin sector 
is a last proof that one can add to the system up to 3 excitations of spin 1 
(Fig.~\ref{spec_s_sqm36_5.00})
which are essentially identical and have extremely weak interactions.

\begin{figure} [h!]
\begin{center}
\resizebox{6cm}{5cm}{\includegraphics{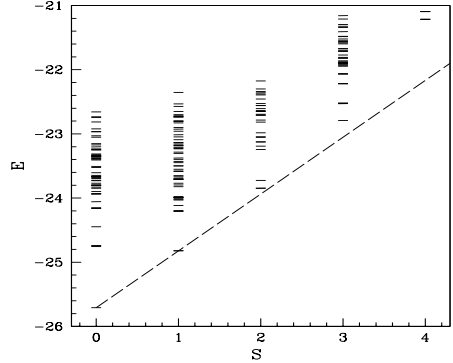}} \end{center}
        \caption[99]{   Same as Fig.~\ref{spec_s_sqm36_5.00}
                     at $J_2/J_1=1.5$.
% The lowest eigenstate belong to the same IR.
        }  \label{spec_s_sqm36_1.50}
\end{figure}

\begin{figure}  [h!]
        \begin{center}
        \resizebox{8cm}{!}{\includegraphics{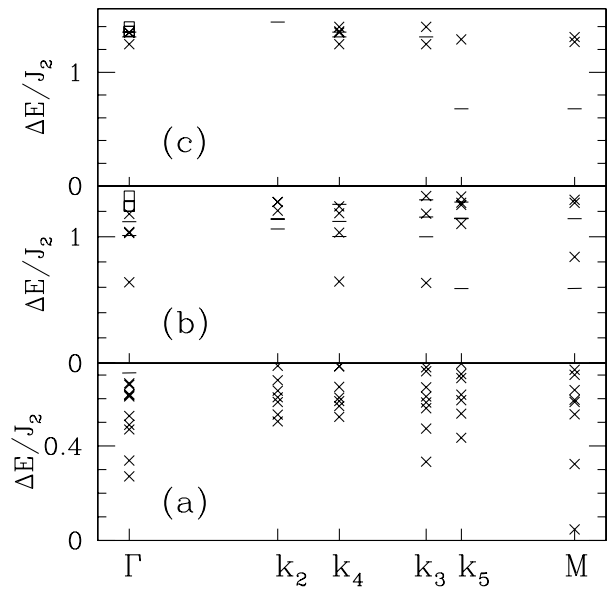}} \end{center}
        \caption[99]{ Lowest excitation energies above
    the ground-state $\Delta E/J_2$ vs wave vector $k$
    for $J_2=1$ (a) $J_2=1.5$ (b) $J_2=5$ (c)
    with spin  $S=0$ (crosses) , $S=1$ (tirets) , $S=2$ (open squares).
        }  \label{fig_band}
\end{figure}

% Something new appears when decreasing the $J_2/J_1$ ratio toward 
% 1.5: the lowest  ``2a-excitations'' merge energetically with the
% 1a-excitations leading to a new picture of the excited states
% as we will show below.

%\begin{figure} [h!]
%        \begin{center}
%        \resizebox{6cm}{!}{\includegraphics{BRIL_sqr.ps}} \end{center}
%        \caption[99]{ Brillouin zone of the square
%lattice. $\Gamma$, X, M, $k_i \; (i=3,5)$ are the five different
%wave vectors present in the N=36 sample.
%        }  \label{brillouin_zone}
%\end{figure}
\begin{table*}  [t!]
\begin{center}
\begin{tabular}{|l l c c r | c l c r r r | c l c r r r |}
| & \multicolumn{4}{c|}{1d} & \multicolumn{5}{c}{$J_2=5$} & | & \multicolumn{5}{c}{$J_2=1.5$} & |\\ \hline
n & $\Delta E$   &$S$& $k$ &$\sigma$& n & $\Delta E/J_2$&$S$&$\bf k$  & $R$ &$\sigma$ &n&$\Delta E/J_2$&$S$ &$\bf k$  & $R$ &$\sigma$ \\ \hline
  &  0.          & 0 &$\pi$&   1    &    &0.           & 0  &$\Gamma$& -1 &  -1     &   &0.           & 0  &$\Gamma$& -1 &  -1 \\ \hline
a &0.684741648   & 1 &  0  &   1    & a  &0.677944871  & 1  & $k_5$  &    &   1     & a &0.590054594  & 1  &  $k_5$ &    &   1 \\
  &              &   &     &        & a  &0.677948816  & 1  & $M$    & $i$&         & a &0.591242101  & 1  &  $M$   & $i$&     \\ \hline
  &              &   &     &        & 2a &1.246710497  & 0  & $k_3$  &    &   1     &2a &0.635033794  & 0  & $k_3$  &    &   1 \\
  &              &   &     &        & 2a &1.246757039  & 0  & $k_4$  &    &   1     &2a &0.640143484  & 0  &$\Gamma$&  1 &   1 \\
  &              &   &     &        & 2a &1.246757966  & 0  &$\Gamma$&  1 &   1     &2a &0.645292084  & 0  & $k_4$  &    &   1 \\ \hline
b &1.302775637   & 0 &  0  &  -1    & b  &1.266511222  & 0  & $M$    & -1 &  -1     & b &0.839974642  & 0  & $M$    & -1 &  -1 \\
  &              &   &     &        & b  &1.288135982  & 0  & $k_5$  &    &  -1     &2a &1.000230994  & 1  & $k_3$  &    &  -1 \\
  &              &   &     &        & b  &1.309791780  & 0  & $M$    &  1 &  -1     &2a &1.001481120  & 1  & $k_4$  &    &   1 \\
  &              &   &     &        & 2a &1.309841802  & 1  & $k_3$  &    &  -1     &2a &1.008659631  & 1  &$\Gamma$& -1 &   1 \\
  &              &   &     &        & 2a &1.309846926  & 1  & $k_4$  &    &   1     &2a &1.029661098  & 0  &$\Gamma$& -1 &  -1 \\
  &              &   &     &        & 2a &1.309863513  & 1  &$\Gamma$& -1 &   1     &2a &1.032523222  & 0  & $k_4$  &    &  -1 \\
  &              &   &     &        & 2a &1.349565139  & 0  &$\Gamma$& -1 &  -1     &2a &1.037137527  & 0  &$\Gamma$&  1 &  -1 \\
c &1.521999231   & 1 &$2\pi/3$&     & 2a &1.349566020  & 0  & $k_4$  &    &  -1     & c &1.060994796  & 1  & $k_2$  &    &  -1 \\
d &1.802775637   & 1 &$\pi/3$&      & 2a &1.349574718  & 0  &$\Gamma$&  1 &  -1     & b &1.100373750  & 0  & $k_5$  &    &  -1 \\
  &              &   &     &        & 2a &1.352732556  & 1  &$\Gamma$&$i$ &         &2a &1.119357730  & 1  &$\Gamma$&$i$ &     \\
  &              &   &     &        & 2a &1.352736474  & 1  & $k_4$  &    &  -1     &2a &1.120391023  & 1  & $k_4$  &    &  -1 \\
  &              &   &       &      & 2a &1.359019058  & 2  &$\Gamma$& -1 &  -1     & c &1.135860677  & 1  & $k_2$  &    &   1 \\

  &              &   &     &        & 2a &1.359020101  & 2  & $k_4$  &    &  -1     & c &1.142923268  & 1  & $k_2$  &    &  -1 \\
  &              &   &     &        & 2a &1.309863513  & 2  &$\Gamma$&  1 &  -1     &3a &1.143097244  & 1  & $k_5$  &    &   1 \\
  &              &   &     &        & 2a &1.399283498  & 2  & $k_3$  &    &   1     &3a &1.143396461  & 1  & $M$    & $i$&     \\
  &2.802775637   & 2 &$\pi$&   1    & 2a &1.399292379  & 2  & $k_4$  &    &   1     & c &1.145902142  & 1  & $k_5$  &    &  -1 \\
  &2.802775637   & 0 &$\pi/3$&      & 2a &1.399293748  & 2  &$\Gamma$&  1 &   1     &2a &1.155611046  & 1  & $k_3$  &    &  -1 \\
\end{tabular}
\end{center}
\caption[99]{ Energy of excitations of the simple Heisenberg
chain of length 6,  of the CCM for $J_1=1$ and $J_2=$
5 (respectively 1.5) on the  $N=36$ sample.
For each eigen-state we display the total spin $S$,
the ${\bf k}$ eigen-vector (as described in Fig.~\ref{checkerboard2}),
the characters of the wave function
respectively in a $\pi/2$ rotation (R) and in a mirror symmetry
($\sigma$). The index n indicates how the excitations of the 2d model
can be build from excitations of the 1d Heisenberg chain.
}
\label{ener_j2_5}
\end{table*}

\begin{figure} [h!]
\begin{center}
\resizebox{6cm}{!}{\includegraphics{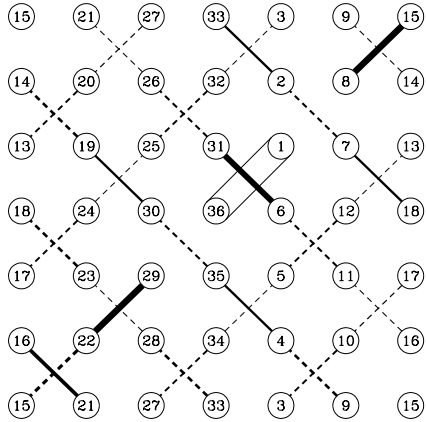}} \end{center}
\caption[99]{  Same as Fig.~\ref{dimer_cor_j2.1.50}
           % Dimer-dimer correlations
         in the 4th excited state of the N=36 sample ($J_2/J_1=1.5$).
        }  \label{dimer-cor-excited}
\end{figure}

Decreasing the ratio $J_2/J_1$, 
% the first $S=1$ excitations have the same properties but the rest of
the spectra testify
of stronger interactions between multi-spinons excitations.
%leading, in particular, to an increase of the band-width of 
In particular,
a new feature does appear: a splitting of the singlet 2a excitations 
with a downward shift in energy of some of them which reach the 
level of the 1a-excitations at $J_2/J_1 = 1.5$
(Table~\ref{ener_j2_5}, Fig.~\ref{spec_s_sqm36_1.50} and Fig.~\ref{fig_band}).
%a striking qualitative new characteristic does appear for this intermediate
%coupling.
%Whereas the $J_2/J_1 = 5$ spectrum can always be quasi quantitatively 
%described by simple additions of excitations of the uncoupled $L=6$
%chains, it is no more the case of the $J_2/J_1 = 1.5$ spectrum.
One can see in Fig.~\ref{fig_band}b this set of singlet states 
with excitation energy $\Delta E \sim 0.6 J_2$   
well separated from the other singlets.
%that disperses weakly toward the corner of the Brillouin zone 
%and more strongly around the line $\Gamma$
%$ \rightarrow (\pi,0)$.
This is a new feature
if we compare it to Fig.~\ref{fig_band}c for $J_2/J_1 = 5$,
where the spectrum is merely that of uncoupled chains.
But it is by no means related to the physics of the
neighboring VBC phase as can be seen by comparing this spectrum
with the VBC spectrum on the same sample Fig.~\ref{fig_band}a.
As explained above the 1a-excitations can be described as purely
"unidimensional",
but these 2a-levels are excitations which cannot be localized
on a single chain.
The large energy gained with inter-chain coupling
suggest some interrelation between chains.
Yet the spin-spin and dimer-dimer correlations in these excitations are
still strongly anisotropic as could be seen in
Tables~\ref{tab-cor-spin-spin-triv},~\ref{tab-cor-dimer-dimer-triv} and
Fig.~\ref{dimer-cor-excited}.
Note that the 2a-excitations occur at wave vectors which can be obtained
by combining wave vectors of the 1a-excitations on the 
sides of the Brillouin zone.
In the thermodynamic limit the 2a excitations are expected to cover
densely the Brillouin zone. 
We conjecture that the appearance of these  excitations
% defined on the whole Brillouin zone
could be the precursor of a collective mode supposed to exist
in a crossed sliding Luttinger liquid \cite{mkl01}.
%This is indeed a very striking feature: energetically  most of the mode
%appears well separated from a quasi continuum which may
%prefigurate the continua of many spinon excitations. This mode
%disperses on the whole 2d Brillouin zone.

\begin{table}  [h!]
\begin{center}
\begin{tabular}{|c|c||c|c|}
  j &   s(ij) &  j  &      s(ij)      \\
\hline
 36 & -0.4044253289 & 4 & -0.0200188490 \\
 15 &  0.1666623287 & 2 & -0.0403867716 \\
 22 & -0.1505704767 & 9 &  0.0164521956 \\ \cline{1-2}
  3 &  0.0505168767 &16 & -0.0129298755 \\ \cline{3-4}
 10 & -0.0536890650 &17 &  0.0523125280 \\ 
 12 & -0.0470765522 &   &               \\ 
\end{tabular}
\end{center}
%36 & -0.4044253289 & 4 & -0.0200188490 \\
%15 &  0.1666623287 & 2 & -0.0403867716 \\
%22 & -0.1505704767 & 9 &  0.0164521956 \\ \hline\hline
% 3 &  0.0505168767 &12 & -0.0470765522 \\
%10 & -0.0536890650 &   &               \\ \hline \hline
%17 &  0.0523125280 &16 & -0.0129298755 \\ \hline \hline
\caption[99]{  Spin-spin correlations in the 4th excited state of 
the spectrum at $J_2/J_1=1.5$.
This state is one of the excited states of the 2-dimensional mode.
As described in  Table~\ref{tab-cor-spin-spin}
it is a singlet state of the trivial RI 
(${\bf k}=(0,0)$, $S=0$, $R=1$, $\sigma=1$).

}
\label{tab-cor-spin-spin-triv}
\end{table}

\begin{table} [h!]
\begin{center}
\begin{tabular}{|c|c||c|c|}
  (k,l) &   D             &  (k,l)  &     D              \\
\hline
  29 22 &  0.0421174442 & 35 30 & -0.0043780206 \\
  22 15 & -0.0099685219 & 11  6 & -0.0043780208 \\
  31  6 &  0.0345888875 &  9  4 & -0.0070999014 \\
  35  4 &  0.0055568217 & 17 10 &  0.0017481736 \\
  28 23 & -0.0011288331 & 34  5 & -0.0004575768 \\
  16 11 & -0.0011288331 & 12  5 & -0.0037470380 \\
  21 16 &  0.0139907353 &       &               \\
\end{tabular}
\end{center}
\caption[99]{ Dimer-dimer correlations in the 4th excited state of the 
$N=36$ sample ($J_2/J_1=1.5$).%Same legends as before.
}
\label{tab-cor-dimer-dimer-triv}
\end{table}

%This seems seems equally
%true on first sight for the spectrum of the $J_2/J_1 = 1.5$ 
%(Fig.~\ref{spec_s_sqm32_1.50}).
%Second, for this range of parameters,
%the low energy spectra calculated for $N=16,32,36$ all remain quite similar
%to the ones found for vanishingly coupled chains:372, ($ J_2/J_1 \gg 1$)
%as may be seen  by comparing the spectrum $J2/J1=1.5$ of the $N=32$ sample
%(see Fig.~\ref{spec_s_sqm36_5.00})
%quite close to the one of uncoupled chains with the spectrum at $J2/J1=1.5$
%(see Fig.~\ref{spec_s_sqm32_1.50}):
%for low values of the total spin $S$, the lowest eigenlevels
%for a given $S$ belong to the same irreducible representations (IR).
%The lowest $S=1$ excitations consist of a set of 4 ($N=16$), 8 ($N=32$),
%or 6 ($N=36$) quasi degenerate states  at all the $k$ vectors
%on the edge of the Brilloun zone ( two states for each $k$)
%separated from the others $S=1$ states (for these finite sized systems).
%The presence of IR at all these $k$ vectors allow the localization
%of excitations along a particular chain.
% These numbers
% of states is equal to the number of chains in the samples.
%Moreover the lowest energies $E(S)$ in each $S$ sector
%increase linearly with $S$ at low $S$
%(up to $S=L$ for large $J_2/J_1$ as seen in Fig.~\ref{spec_s_sqm36_5.00}
%and limited to lower $S$ value at $J_2/J_1=1.5$
%as seen in Fig.~\ref{spec_s_sqm32_1.50}).
%A natural explanation arises from the 1d picture, arguing that these
%low eigenstates correspond to the addition of $S=1$ excitations
%on each individual chain.

All these results  (spectra, correlations, scaling of the spin gap) 
indicate a 1d behavior which
extend down to $J_2/J_1=1.5$,
i.e. well beyond the weak coupling limit investigated by (SSL).
The 2-dimensional S=0 excitation mode singled out
for $J_2/J_1=1.5$ on the N=36 sample may be an illustration of
the collective mode of the `` crossed sliding Luttinger liquid'': it is
highly anisotropic as could be expected. Its detailed properties
should nevertheless be taken with some care due to the numerous
constraints  associated with periodic boundary conditions on
 these small samples.

The precise location of the lower limit of quasi 1d behavior
is clearly out of reach of exact ED calculations. Nevertheless
major modifications of the spectra when $J_2/J_1$ varies are a
good qualitative indication of this transition.
% The
%transition appears rather abrupt (see Figs.~\ref{energies},~\ref{spin_gaps}) 
%but this may be an artefact of the small sizes.

At the isotropic point $J_2/J_1=1$ the ground-state is a VBC 
with LRO in 4-spin S=0 
plaquettes\cite{fmsl01}:
this phase does not break $SU(2)$ symmetry (it has a spin gap),
 but breaks the space symmetry of the lattice. This last feature
manifests itself 
by the existence of a very low lying $S=0$ excited state which
collapses to the ground-state in the thermodynamic limit. The
absolute ground-state belong to the trivial irreducible
representation of the space group (wave vector $k=(0,0)$, even in all
operations of $C_{4v}$) and the first excited S=0
state, which embodies the VBC symmetry breaking
is a state with wave-vector $(\pi,\pi)$, odd under a $\pi/2$ rotation
around O  (These two states form the doublet called ${\cal
E}_{VBC}$ in the following). 
When $J_2/J_1$ increases beyond 1, the first manifestation of the
instability of the VBC  phase  appears in
the modification of the symmetries of these low lying singlets,
which is concomitant with the appearance of quasi
1-dimensional excitations in the S=1 sector: these two
manifestations appear as soon as  $J_2/J_1= 1.05$ (resp 1.10) for
$N=32$ (resp 36). The transition shift to larger $J_2/J_1$ values
with increasing sizes. Its location in the thermodynamic limit is
difficult with present sizes.
%Between $J_2/J_1\sim 1.10$ 
%and $J_2/J_1 \ge 1.25$ the features of the spectra are extremely involved and
%probably extremely sensitive to finite size effects. 
At $J_2/J_1= 1.5$, as seen above, 
all characteristics (spectra and correlations functions) 
point to a sliding Luttinger liquid.
At $J_2/J_1=1.25$, the scaling of the spin-gap (and the spectra)
also suggest this behavior.
The transition might occurs just below this point.
%but there are still some singlets
%below the first excitation in the S=1 channel which obscure the
%picture: this may be a finite size effect (as the numerous singlets
%in the VBC spectra are) but the size and lengths available do not
%allow us to conclude  definitely on this point.  

\subsection{Coexistence of N\'eel LRO and VBC order}

We now turn to the investigation of the extension of 
the N\'eel and the VBC phases and the question of their possible
coexistence.

\begin{figure} [h]
        \begin{center}
        \resizebox{6cm}{5cm}{\includegraphics{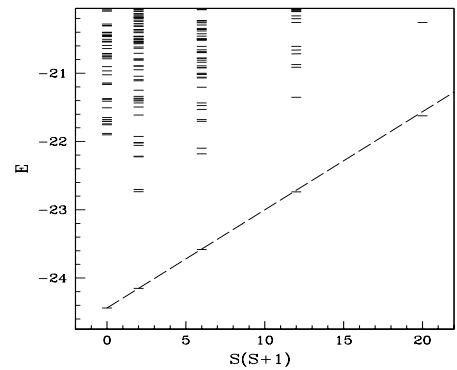}} \end{center}
        \caption[99]{ Spectrum at $J_2=0$ (Heisenberg model on the
square lattice) vs $S(S+1)$ for $N=36$.
% The dashed line joins the lowest $S=0$ and $S=1$ states.
The QDJS characteristic of collinear N\'eel LRO
at the bottom of each $S$ are well aligned (dashed line)
and clearly separated from the other lowest excitations (see text).
        }  \label{spec_sqr_heis36}
\end{figure}

As discussed in \cite{lblps97,lmsl00} 
the simplest way to monitor the extension
of N\'eel LRO is to search for the
so-called quasi degenerated joint states (QDJS)
which allow to break $SU(2)$ and lattice space symmetry 
in the thermodynamic limit.
This is a set  of low lying states appearing  in each total spin sector $S$
up to $S \approx \sqrt{N}$, belonging to specific IRs of the
space group, 
which collapse into a degenerate ground-state as $1/N$ 
for $N\rightarrow\infty$ and evolve as $S(S+1)$ with $S$.
For collinear N\'eel LRO there is one QDJS for every $S$.
Presently these states belong  to the trivial IR of the space group
 for $S$ even and to
the $k=(0,0)$ IR, even under  a mirror symmetry  and odd under a $\pi/2$
rotation representation, when S is odd. 
In a plot of the spectrum vs $S(S+1)$ the QDJS form a line 
well separated
from the macroscopic excitations (magnons) which only collapse as
$1/\sqrt{N}$ as shown in  Fig.~\ref{spec_sqr_heis36} 
for $J_2=0$.
The $\sim 1/N$ collapse of the QDJS implies that  the spin-gap $\Delta$
also closes  as $\sim 1/N$.

On the other hand 
the beginning of the plaquette phase may be estimated
by the appearance of the  quasi degenerate S=0 doublet 
(${\cal E}_{VBC}$) 
sketched in the previous subsection.

The destruction of the QDJS and the appearance of ${\cal E}_{VBC}$ are
not necessarily simultaneous.
Both the N\'eel and the VBC phase may in principle
co-exist in a  gapless phase.
In a recent paper Sachdev and Park \cite{sp01}, following
earlier investigations in Ref~\cite{sow01},
have suggested  that the destabilization of collinear N\'eel LRO
by 2th order transition into a VBC phase must generically occur
within such a coexistence region.

Linear spin-wave calculations \cite{ssf98,c01} predict that
N\'eel LRO survive quantum fluctuations till $J_2/J_1 \approx 0.75$.
Up to this value the characteristic features of N\'eel LRO 
are rather well observed 
in the spectra and in the evolution of $\Delta$ and $E/N$ with $N$. 
Well below $J_2/J_1=0.75$, however, one may notice
a slight curvature of the line of QDJS 
and there is small irregularity
in the evolution of $\Delta$ and $E/N$ with $N$:
$\Delta$ is slightly larger and $E/N$ slightly lower for $N=36$ than
for $N=32$ 
as might be seen in Fig.~\ref{spin_gaps} and Fig.~\ref{energies}.
Nevertheless $\Delta$ seems to extrapolate rather well
to zero at $N\rightarrow\infty$ at $J_2/J_1=0.65$.
And for $J_2/J_1=0.75$ the $1/N$ extrapolation of the $N=16,32,36$  
values of $\Delta$ point to a spin-gap at most very small,
$\sim 0.06$,
a value of the order of
the corrections to the asymptotic regime  which are expected to be large
and negative as we approach the boundary of the N\'eel 
phase~\cite{hn93,fsl01}. 
So the N\'eel phase might persist up to this point.
On the other hand, up to $J_2/J_1=0.75$, 
the separation of the two ${\cal E}_{VBC}$
states appears likely to remain finite as $N\rightarrow\infty$.
Thus the VBC phase is not reached.

\begin{figure}
        \begin{center}
        \resizebox{6cm}{5cm}{\includegraphics{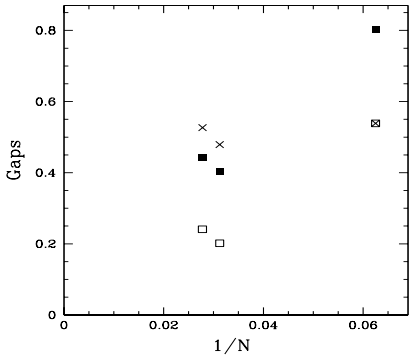}} \end{center}
        \caption[99]{ Spin gaps $\Delta$ (full squares),
    separation of the two singlet states of ${\cal E}_{VBC}$ (open squares),
    gap between the ground-state and the third singlet (crosses)
    at $J_2/J_1=0.85$.
   % The dashed line is a fit to the $\Delta$ values.
        }  \label{gap_sqm_0.85}
\end{figure}

By contrast results at $J_2/J_1=0.85$ point to a finite spin gap 
and a degeneracy of the two states of ${\cal E}_{VBC}$
which remain separated from the other singlet states 
for $N\rightarrow\infty$ (Fig.~\ref{gap_sqm_0.85}).

These results lead to conclude to a transition between the 
two phases close to $J_2/J_1 \approx 0.75$ in agreement with 
the spin-waves result and that a coexistence region of 
the two phases either  occurs at most in a very small range
or is absent.

\section{Summary}

Exact diagonalization have allowed to characterize 
three phases of the CCM: the $(\pi,\pi)$ N\'eel phase at small
diagonal couplings, 4-spin S=0 plaquette VBC at the isotropic
point, and sliding Luttinger liquid for $J_2/J_1 \ge 1.25$.
The transition between the N\'eel phase  and the VBC phase
occurs at $J_2/J_1 \approx 0.75$. No evidence of a coexistence region 
predicted by Sachdev and Park \cite{sp01} is found.
The CCM exhibits 1d behavior for a large
range of parameters ($J_2/J_1$ greater than $\approx 1.25$)
where the transverse coupling between the
chains is of the order of the intra-chain coupling.
This 1d behavior is likely to be truly  
realizing the  $SU(2)$ "Luttinger sliding phase"
with deconfined spinons predicted by Starykh, Singh and Levine~\cite{ssl02}.

Acknowledgments: We acknowledge fruitful discussions with 
B. Canals, C. Lacroix, P.  Lecheminant, O.
Starykh, R.R.P. Singh and A. Vishwanath.
Computations were performed at The Centre de Calcul pour la Recherche de
l'Universit\'e Pierre et Marie Curie and at the Institut de
d\'eveloppement des Recherches en Informatique Scientifique du
CNRS under contract 990076.

%\bibliography{claire}

\end{document}